\documentclass[
reprint,
groupedaddress,
showpacs,
showkeys,
amsmath, amssymb,
aps,
prl,
]{revtex4-2}

\usepackage{xcolor}
\usepackage{graphicx}
\usepackage{dcolumn}
\usepackage{bm}
\usepackage[colorlinks=true, allcolors=blue]{hyperref}
\usepackage{comment}
\usepackage[normalem]{ulem}
\graphicspath{{figures/}}
\usepackage[separate-uncertainty=false]{siunitx}
\usepackage{braket}
\usepackage{xspace}
\usepackage{xstring}
\newcommand{\op}[1]{\widehat{#1}}
\newcommand{\ham}{\op{\mathcal H}}
\newcommand{\na}{$^{23}$Na\xspace}

\newcommand{\blue}[1]{\textcolor{blue}{}}
\newcommand{\suppress}[1]{\textcolor{brown}{}}

\newcommand*{\aref}[1]{%
	\IfBeginWith{#1}{eq:}{Eq.~\eqref{#1}}{}
	\IfBeginWith{#1}{fig:}{Fig.~\ref{#1}}{}%
	\IfBeginWith{#1}{tab:}{Table~\ref{#1}}{}%
	\IfBeginWith{#1}{appendix:}{Appendix~\ref{#1}}{}%
	\IfBeginWith{#1}{sec:}{Section~\ref{#1}}{}%
	}

\makeatletter
\def\maketitle{
\@author@finish
\title@column\titleblock@produce
\suppressfloats[t]}
\makeatother

\begin{document}

\title{Observation of Massless and Massive Collective Excitations with Faraday Patterns in a Two-Component Superfluid}
\date{\today}
\author{R. Cominotti}
\thanks{These two authors contributed equally to this work.}
\author{A. Berti}
\thanks{These two authors contributed equally to this work.}
\author{A. Farolfi}
\author{A. Zenesini}
\email[]{alessandro.zenesini@ino.cnr.it}
\author{G. Lamporesi}
\email[]{giacomo.lamporesi@ino.cnr.it}
\author{I. Carusotto}
\author{A. Recati}
\email[]{alessio.recati@ino.cnr.it}
\author{G. Ferrari}

\affiliation{INO-CNR BEC Center and Dipartimento di Fisica, Universit\`a di Trento, and Trento Institute for Fundamental Physics and Applications, INFN, 38123 Povo, Italy. }

\begin{abstract}
We report on the experimental measurement of the dispersion relation of the density and spin collective excitation modes in an elongated two-component superfluid of ultracold bosonic atoms. Our parametric spectroscopic technique is based on the external modulation of the transverse confinement frequency, leading to the formation of density and spin Faraday waves. We show that the application of a coherent coupling between the two components reduces the phase symmetry and gives a finite mass to the spin modes. 
\end{abstract}

\maketitle

The concept of collective excitations is a cornerstone for our understanding of the physics of condensed matter systems. In particular, arguments based on their dispersion relation have provided first insight on the microscopic origin of superfluidity in liquid Helium~\cite{Pines96}. In the specific case of dilute atomic Bose-Einstein condensates (BEC), the Bogoljubov theory, based on a linearized quantum theory around a condensate, provides quantitative predictions for the dispersion relation, with a gapless (massless) sonic behaviour at small wave vectors  followed by a quadratic single-particle one at larger wave vectors~\cite{Pitaevskii16}.

The situation gets more interesting in the case of two-component superfluids, whose collective excitation spectrum consists of two gapless branches for the spontaneously broken U(1)$\times$U(1) symmetry due to the  conservation of particle number in each component. For equal masses and interaction constants of the two components, the two branches are associated to oscillations of the total density or of the density difference, the so-called {\em density} and {\em spin} modes~\cite{Abad2013}. 
At low $k$, both branches have a linear dispersion, yet with generally distinct values of the speed of sound. If the particle number in each component is not conserved, e.g. by applying a field that coherently couples the two components, exciting the condensate relative phase requires an energy cost. As a result, while the massless nature of the total density mode is protected by the Goldstone theorem associated to the remaining U(1) symmetry, the spin mode acquires a finite mass~\cite{Goldstein97}.

Precise information on the Bogoljubov dispersion of single-component condensates was extracted using Bragg spectroscopy~\cite{Steinhauer02}. 
Parametric excitation of a superfluid was pioneered  using a time-dependent modulation of the optical lattice depth~\cite{Schori2004,Stoferle04,Kraemer05,Tozzo05}, by acting on the transverse potential of an elongated harmonically trapped BEC~\cite{Engels2007,Jaskula2012,Smits18} or through a modulation of the interaction constant~\cite{Pollack10,Hung2013, Nguyen2019}.
Distinct spin and density sound velocities were measured in a two-component sodium system by locally perturbing the system with spin sensitive or insensitive potential~\cite{Kim20}.
Two-dimensional bosonic superfluids~\cite{Zhang2020} and strongly interacting superfluids~\cite{Patel20,Hernandez2021} were also recently investigated.

\begin{figure}[b]
    \centering
    \includegraphics[width = \columnwidth]{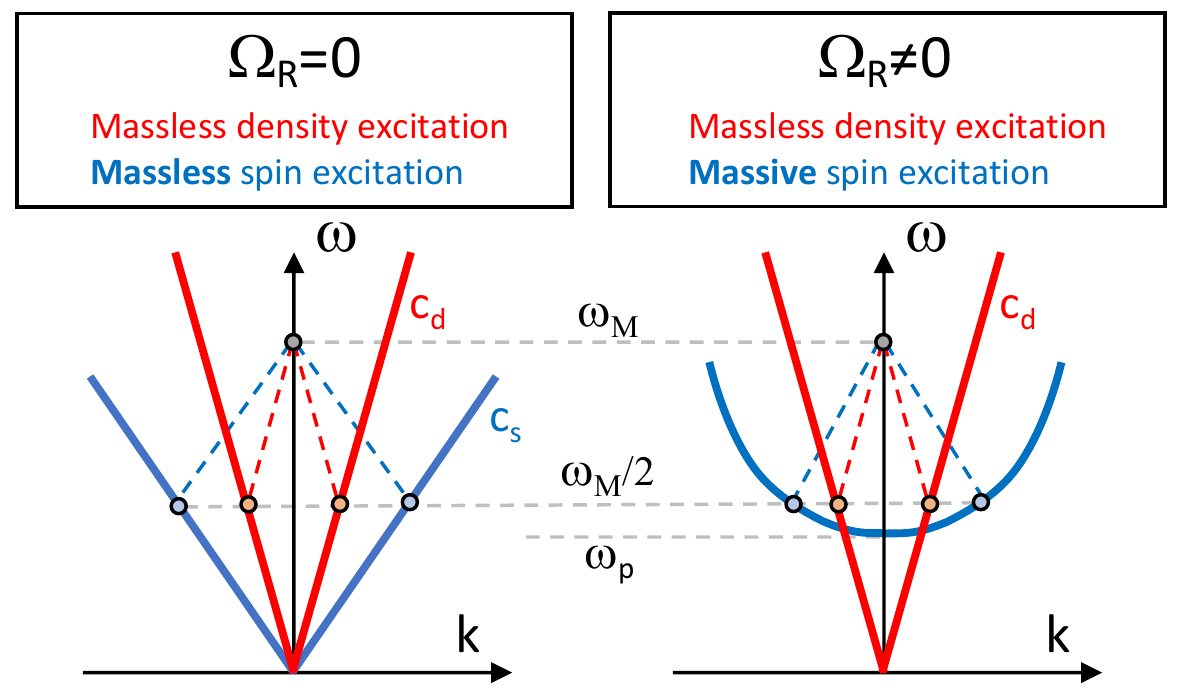}
    \caption{(a) Generation mechanism of excitation pairs in the density and spin branches of a two-component system.
    The external excitation at $\omega_M$ is converted into two excitations with opposite wave vector and half the energy. 
    In the absence of coupling between the two components, $\Omega_R=0$ (left), two symmetries are preserved and both modes have a linear behaviour, with density speed of sound $c_d$ and spin speed of sound  $c_s$. When a coherent coupling is present, $\Omega_R\neq 0$  (right), one of the two symmetries is broken, introducing a curvature in the spin dispersion relation, which makes the excitation acquire a mass.}
    \label{fig:fig1}
\end{figure}

In this Letter, we apply the parametric excitation technique to the novel case of a two-component BEC of ultracold sodium atoms in two spin states. A well-controlled generation of Faraday waves in both the density and in the spin channel allows us to perform a first quantitative and complete measurement of the dispersion relation of the two branches of collective density and spin excitations. 

\begin{figure}[t]
    \centering
    \includegraphics[width = \columnwidth]{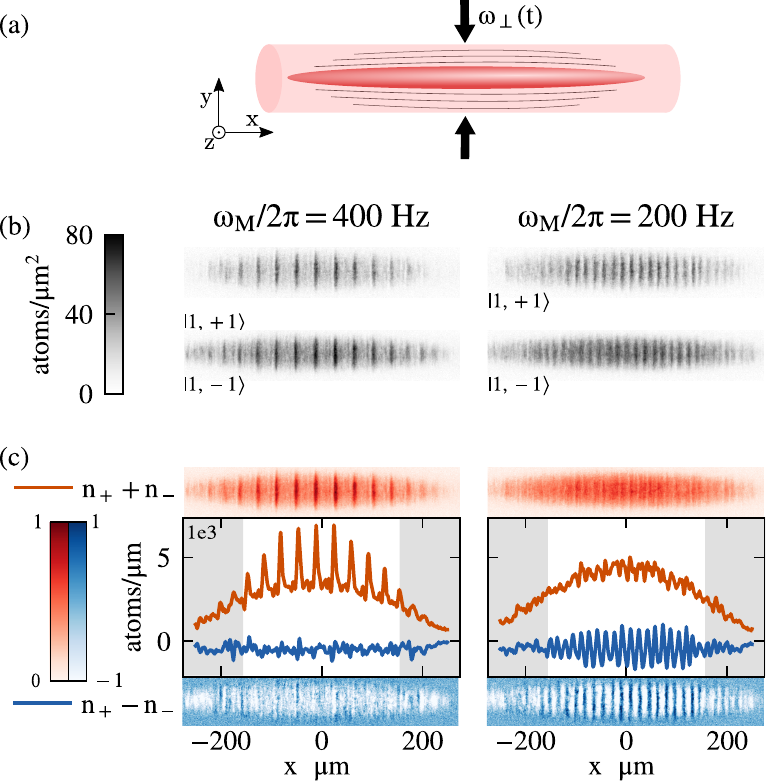}
    \caption{Density and spin Faraday patterns.
    (a) Sketch of the experimental configuration. The transverse trapping frequency is modulated in time at frequency $\omega_M$, periodically compressing the elongated condensate.
    (b) The parametrically generated excitations appear as a spatial pattern with a well-defined wavelength along the axis of the condensate.
    (c) Density (red) and spin (blue) 2D experimental patterns and corresponding integrated 1D profiles for $\omega_M / 2\pi = \SI{400}{Hz}$ (left) and
    $\omega_M/2\pi = \SI{200}{Hz}$ (right). }
    \label{fig:fig2}
\end{figure}

In an intuitive way, one can understand the parametric excitation process as the emission of a pair of phonons (of frequency $\omega_M/2$ and opposite wave vectors $\pm k$) by some classical external drive at $\omega_M$, as sketched in \aref{fig:fig1}. The two modes lead to a spatial pattern oscillating in time, known as Faraday wave [see \aref{fig:fig2}(b)].  Its spatial periodicity is $2\pi/k$ and its visibility oscillates in time at $\omega_M$ (see, e.g., \cite{Nicolin2007}). 

As pictorially shown in \aref{fig:fig1}, the parametric process is active on both density and spin channels (hereafter labeled as $d$ and $s$).
In particular, energy-momentum conservation predicts different values for the wave vector $k$ of the emitted density and spin excitations. 
Concretely, for a coherently-coupled BEC of atoms with mass $m$, the dispersion relations read (see \cite{Abad2013} and references therein)
\begin{eqnarray}
    \omega_d(k) &=& \sqrt{\frac{\hbar k^2}{2m}\left(  \frac{\hbar k^2}{2m} + \frac{2\mu_d}{\hbar} \right)}
    \label{eq:omegaD} \\
    \omega_s(k) &=& \sqrt{\left(\frac{\hbar k^2}{2m} + \Omega_R\right)\left( \frac{\hbar k^2}{2m} + \frac{ 2\mu_s}{\hbar} + \Omega_R\right)}.
    \label{eq:omegaS}
\end{eqnarray}
Here  $\mu_{d,s} = (g\pm g_{12})n_{\text{eff},d,s}/2$ are the effective chemical potentials for the density and spin channels, where $g$ and $g_{12}$ are the intra- and intercomponent interaction constants, and $n_{\text{eff},d}=n_0/2$ and $n_{\text{eff},s}=2n_0/3$, the effective densities after properly including the effects of the geometrical reduction \cite{SM}. 
The strength of the coherent coupling -- that breaks the relative atom number conservation -- is given by the Rabi frequency $\Omega_R$. 
For small $k$ and $\Omega_R=0$, both channels, \aref{eq:omegaD} and \aref{eq:omegaS}, show the sonic-like behaviour $\omega_{d,s}(k)\simeq c_{d,s} |k|$ with speeds of sound $c_{d,s}=\sqrt{\mu_{d,s}/m} $. 

The coherent coupling $\Omega_R$ has no effect on the density branch, while a frequency gap $\omega_p=\sqrt{\Omega_R(\Omega_R+ 2 \mu_s/\hbar)}$ opens in the spin branch, which then turns massive, $\omega_s(k) \simeq \omega_p+\hbar k^2/(2M)$, with an effective mass $M = 2 m \omega_p \Omega_R/(\omega_p^2+ \Omega_R^2)$. 
As already mentioned, the presence of a gap at $k=0$ in the spin channel originates from the explicit breaking of the continuous symmetry U(1), related to the conservation of the relative particle number. For non-interacting atoms the gap is simply $\Omega_R$ and corresponds to the energy cost to slightly move the ground state from its position, i.e., the cost to modify the relative phase of the two hyperfine wave functions locked by an external drive. In the presence of many-body interactions, the problems becomes equivalent to the so-called internal Josephson effect \cite{Williams99,Zibold10} and $\omega_p$ corresponds to the frequency of small oscillations around the homogeneous ground state, which for this reason is also called plasma-frequency (see the recent review \cite{Recati22} and reference therein).

We start our experiments by preparing a BEC of $10^6$ \na atoms in the $\ket{F,m_F}=\ket{1,-1}$ internal state, $F$ being the total atomic angular momentum and $m_F$ its projection on the quantization axis, set by a uniform magnetic field \cite{Colzi18}.
The BEC (with negligible thermal component) is held in a cylindrically symmetric single-beam optical trap  with trapping frequencies  $\omega_{\perp} / 2\pi = \SI{1}{kHz}$ and $\omega_{x} / 2 \pi=\SI{10 }{Hz}$, leading to a Thomas-Fermi profile with radii $r_{\perp}= \SI{3} {\mu m}$ and $r_x= \SI{300}{\mu m}$, for the transverse and longitudinal directions, respectively. 

\begin{figure}[t]
    \centering
    \includegraphics[width = \columnwidth]{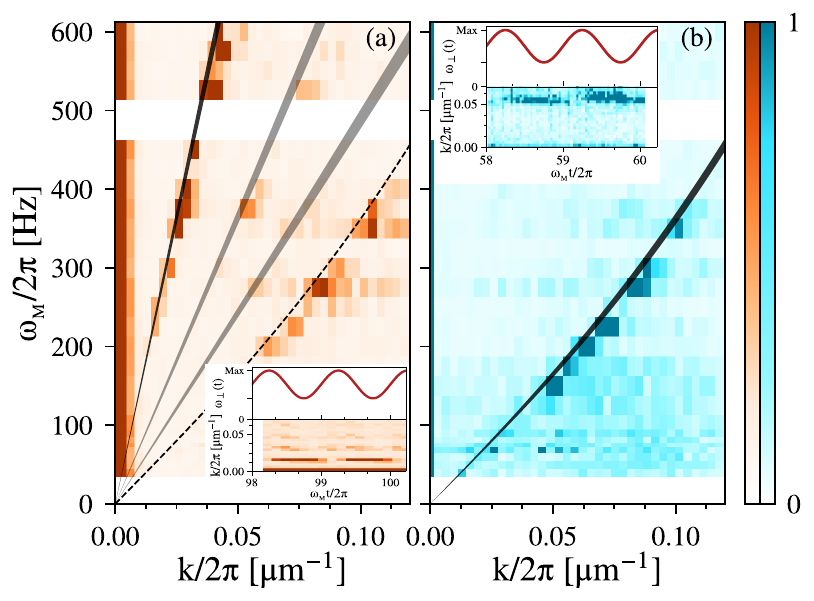}
    \caption{PSD of density (a) and spin  (b) excitations as a function of the modulation frequency. The thick lines indicate theoretical predictions (\ref{eq:omegaD}-\ref{eq:omegaS}) for the dispersion relations (dark) and sub-harmonics (light) (see text), with $\Omega_R=0$ and no fitting parameters. The line thickness corresponds to one standard deviation confidence interval originating from the uncertainty in the atomic density. Insets show the modulation amplitude in time and the corresponding fringe visibility for $\omega_M / 2\pi = \SI{200}{Hz}$. The dashed line in panel (a) indicates the position of the spin branch, where a spurious signal, due to the cross-talk between spin and density modes, is present.
    }
    \label{fig:fig3}
\end{figure}

\begin{figure}[t]
    \centering
    \includegraphics[width = \columnwidth]{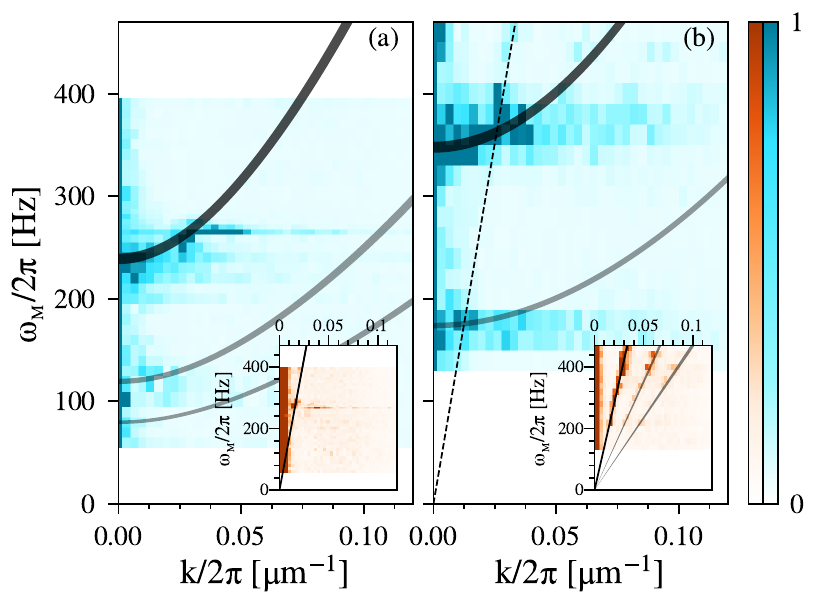}
    \caption{PSD of the spin excitations in the presence of a coherent coupling. In (a)
    $\omega_p/2\pi = 120$ Hz, $\Omega_R/2\pi=33$\,Hz and  $\mu_s/h=225$\,Hz, in (b)
    $\omega_p/2\pi = 175$ Hz, $\Omega_R/2\pi=80$\,Hz and $\mu_s/h=150$\,Hz. The thick lines indicate the  theoretical predictions (\ref{eq:omegaD}-\ref{eq:omegaS}) for the dispersion relations (dark) and sub-harmonics (light) (see text). The line thickness corresponds to one standard deviation confidence interval originating from the uncertainty in the atomic density. Insets show the corresponding PSD of the density channel (unaffected by the coupling). The dashed line in panel (b) indicates the position of the density branch, where a spurious signal, due to the cross-talk between spin and density modes, is present. 
    }
    \label{fig:fig4}
\end{figure}

The two-component BEC is then prepared through an adiabatic rapid passage (ARP) sequence~\cite{Farolfi20,Farolfi21}, which coherently transfers half of the atomic population to the $\ket{1,1}$ state, using a two-photon microwave transition \cite{SM}.
At the end of the ARP, the microwave drive is either completely switched off (experiments in Fig.\,\ref{fig:fig2}-\ref{fig:fig3}) or kept on at the desired value of the coherent coupling between the two components (experiments in \aref{fig:fig4}).

As done in Ref.~\cite{Engels2007}, we induce Faraday waves by
modulating the transverse trapping frequency as $\omega_{\perp}(t)=\omega_{\perp}(0)[1+\alpha\sin(\omega_{M}t)]$, with frequency $\omega_M$ and amplitude $\alpha \in [0.38-0.6]$ [\aref{fig:fig2}(a)]. 
The modulation is applied for a time $t \in [50-400]\,\SI{}{ms}$. 
Since $\omega_x \ll \omega_M < \omega_{\perp}(t)$, the transverse size adiabatically changes in time following the periodic compression and decompression of the potential. In this way no transverse excitation is generated. Conversely, axial modes can be excited, leading to longitudinal (1D) Faraday waves \cite{Modugno06,Nicolin2007}.
At the end of the modulation, the trapping potential is suddenly removed and the atoms in the $\ket{1,-1}$ ($\ket{1, +1}$) state are selectively imaged after a short time of flight (TOF) of \SI{2}{ms} (\SI{3}{ms}) [\aref{fig:fig2}(b)]. Due to the short duration of the TOF stage, our very elongated condensate expands only in the transverse directions, leaving the axial distribution practically unchanged.
We can therefore integrate the absorption images displayed in \aref{fig:fig2}(b) over the transverse directions to extract the 1D densities $n_\pm$ in the  $\ket{1,\pm 1}$ spin states and, from these, the total density $(n_+ +n_-)$  and spin $(n_+-n_-)$ profiles along $x$. In \aref{fig:fig2}(c), we present typical profiles in the absence of coherent coupling for two different values of the modulation frequency. Depending on the modulation frequency, we observe that a periodic pattern can be formed  in the density (left) or in the spin (right) profiles.

The strength and periodicity of the spatial modulations can be quantified by calculating the power spectral density (PSD) of the 1D profiles as:
\begin{equation}
   \text{PSD}_{d, s} (k)= \left|\int{(n_+ \pm n_-) e^{i k x} \,\mathrm{d}x } \right|^2.
   \label{PSD}
\end{equation}
To suppress inhomogeneous broadening effects, we restrict the analysis to the central \SI{300}{\um} of the condensate [white region in \aref{fig:fig2}(c)].

Examples of the time-evolution of the PSD are shown in the insets of
\aref{fig:fig3}(a-b). The PSD displays periodic oscillations at specific values of $k$, with the same frequency of the modulation. This behaviour is typical of Faraday waves and in close agreement with the theoretical predictions for a single-component condensate \cite{Modugno06, Nicolin2007}, the visibility of both spin and density excitations is maximal in time when  $\omega_{\perp}(t)$ is minimum [$ t=(2n+3/2)\pi/\omega_M$ with $n \in \mathbb{Z}$]. Being spin and density excitations always in phase, the ratio between the two visibilities remains constant within a modulation period. While the position of the PSD peak is stable, its absolute height is strongly dependent on the chosen values of the modulation strength and duration. In order to optimize the visibility of the features for all the modulation frequencies, the duration or amplitude of the modulation is empirically optimized and the atoms are released always at times corresponding to a minimum in $\omega_\perp(t)$ (see insets in \aref{fig:fig3}).
This protocol allows us to obtain the PSD as a function of the modulation frequency $\omega_M$ and the wave vector $k$ of the induced pattern reported as color plots in \aref{fig:fig3} and in \aref{fig:fig4}.

The values of $k$ for which the pattern is strongest follow extremely well the theoretically predicted dispersion relations Eqs.\,(\ref{eq:omegaD}-\ref{eq:omegaS}) with $\Omega_R=0$, as reported in \aref{fig:fig3}.  Here, the only non-trivial parameter is the BEC peak density $n_0$ of the 3D distribution, which is independently calibrated by measuring the plasma frequency of the two-component BEC at the center of the cloud~\cite{SM, Farolfi21}, leading to the estimated chemical potentials, $\mu_d/h=\SI{3}{kHz}$  and $\mu_s/h= \SI{145}{Hz}$.

In the density channel, the $k$ position of the peak depends linearly on the modulation frequency as $k\simeq \omega_M/ (2 c_d)$, since all the probed frequencies are well in the sonic region of the dispersion relation, $\omega_M \ll \mu_d/\hbar$.
In contrast, in the spin channel the linear dependence $k\simeq \omega_M/(2c_s)$ is restricted to modulation frequencies $\omega_M \lesssim 2\mu_s/\hbar$, whereas for larger $\omega_M$, a deviation from the linear behaviour is observed, in agreement with the supersonic nature of the Bogoljubov dispersion.

We notice that a residual signature of the spin modes is visible on the density spectra and vice versa [see dashed lines in \aref{fig:fig3}(a) and  \aref{fig:fig4}(b)] This originates from a weak coupling between the spin and density modes due to slight imbalance of the density in the two spin states, as well as from some cross-talk between the two spin components in the imaging technique~\cite{Farolfi21}.

We turn now to the analogous measurement performed in the presence of the coherent coupling, $\Omega_R\neq 0$, shown in \aref{fig:fig4}. 
In order to observe collective behaviors in the spin channel, we need to have $\hbar \Omega_R < 2 \mu_s $, a condition which ensures also that the adiabaticity condition is fulfilled, having $\omega_p\ll \omega_\perp$. Operating in this regime is experimentally possible thanks to the $\mu$G-level stability in the magnetic field that is provided by the magnetic shielding surrounding our setup \cite{Farolfi19, SM}. This  magnetic field stability corresponds to a frequency stability of the atomic resonance at the level of a few Hz.

The results for $\Omega_R\ne 0$ are discussed in the following, with peculiar attention to: 
{\it i}) the \textit{massive dispersion} in the spin channel, {\it ii}) the \textit{sonic dispersion} in the density channel, unaffected by the coherent coupling; {\it iii}) the \textit{scaling} of the gap with $\omega_p$. \\
We verified, both numerically and experimentally, that the formation of the spin pattern occurs on much faster time-scales, as compared to the case $\Omega_R=0$, when the modulation frequency $\omega_M$ is close to $2\omega_p$. 
Therefore, in order to access all the above mentioned points, different values of the amplitude and modulation time were used for the data sets shown in \aref{fig:fig4}(a,b).

To address point \textit{i}), we use short modulation times, that allow to obtain a clean signal in the spin channel around $\omega_M \simeq 2\omega_p$, showing the massive character of spin collective excitations [\aref{fig:fig4}(a)]. This choice inhibits the observation of the dispersion relation in the density channel (see the inset). To overcome such a difficulty, we increase depth and duration of the periodic driving and the corresponding measurement is shown in \aref{fig:fig4}(b): the inset reveals how the excited wavevector $k$ still depends linearly on $\omega_M$, regardless of the coherent coupling [point \textit{ii})]. 
Unfortunately, at such long modulation times, spin modes around the plasma frequency are excited strongly in the non-perturbative regime, which results in the broadening of the spectrum in $k$-space: hence, the blue horizontal stripes in \aref{fig:fig4}(b) should not be confused with the expected dispersion relation Eq.\,(\ref{eq:omegaS}).\\
For what concerns point \textit{iii}), the different values of $\Omega_R/2\pi=\SI{33}{Hz}$ and $\Omega_R/2\pi = \SI{80}{Hz}$ in the data sets of panels (a) and (b) of \aref{fig:fig4} illustrate the scaling of the gap size with $\omega_p$. In the spin channel an appreciable signal is only visible above $2\omega_p$. Above this point, the $k$ position of the peak remains in very good agreement with the expected massive dispersion relation of the spin excitations (continuous black line). For the data in \aref{fig:fig4}(a) and \aref{fig:fig4}(b), the effective mass is about 25\% and 75\% of the atomic mass, respectively.

While the dispersion relations \aref{eq:omegaD} and \aref{eq:omegaS} are consistent with the dominant signal in the PSD of \aref{fig:fig3} and \aref{fig:fig4}, the additional signals at sub harmonic frequencies (grey lines) are typical of Faraday instabilities described in terms of the Mathieu equation \cite{SM}, which is known to show instability (resonance) regions around $\omega_M=2 \omega(k)/l$, with $l$ a positive integer number \cite{Landau1935}. We verified that such signals are not originated by residual anharmonicities of the optical trap modulation.

In conclusion, in this work we have made use of a parametric excitation technique to perform a detailed measurement of the dispersion relation of the longitudinal density and spin collective excitations of an elongated two-component BEC. The accuracy and flexibility of our spectroscopic technique directly hint at its application to more complex phenomena in quantum mixtures. Specifically, in the case of an interspecies interaction larger than the intraspecies one, we could study spin excitations when  crossing the  ferromagnetic phase (see~\cite{Abad2013} and reference therein). Another relevant application of our technique would be the study of the dispersion relations in the various phases of spin-orbit coupled mixtures. Particularly interesting is the rotonization of the spectrum and the appearence of a new Goldstone mode in the so-called stripe phase (see \cite{Martone15} and reference therein).

It is also worth mentioning that the parametric excitation of the spin modes in our system is essentially equivalent to the so-called parallel pumping amplification, which has been taking a very important role in magnonics \cite{Bracher17}. In this respect, a byproduct of the present work, is to provide a further evidence that our platform, as  already shown in \cite{Farolfi21b}, provides new insights in the dynamics of magnetic materials.

From a yet different perspective, analog models~\cite{Barcelo2011} based on two-component atomic BECs are a promising platform for quantitative studies of quantum field theories on curved space-times, such as cosmological particle creation and analog Hawking radiation~\cite{Fischer2004,Visser2005,Butera2021,Butera2017}. In particular, the control on the mass of the spin excitations, that was demonstrated here via coherent coupling, is of great interest in view of extending this research to the case of massive fields interacting with the gravitational background. On a longer run, the complex dynamics that is obtained when the externally-imposed modulation of the trap parameters is replaced with an excitation of the transverse degrees of freedom of the BEC, may provide information on back-reaction phenomena of the quantum field theory on the background space-time~\cite{Butera2021b}.\\

\begin{acknowledgments}
We thank F. Dalfovo and S. Stringari for fruitful discussions and for their critical reading of the manuscript. IC acknowledges continuous collaboration with S. G. Butera. We acknowledge funding from Provincia Autonoma di Trento, from INFN through the FISH project and from the Italian MIUR under the PRIN2017
project CEnTraL (Protocol Number 20172H2SC4). 
This work was supported by Q@TN, the joint lab between University of Trento, FBK - Fondazione Bruno Kessler, INFN - National Institute for Nuclear Physics and CNR - National Research Council.
\end{acknowledgments}

\clearpage

\widetext
\begin{center}
\textbf{\large Supplemental Materials: Observation of Massless and Massive Collective Excitations with Faraday Patterns in a Two-Component Superfluid}
\end{center}

\twocolumngrid

\setcounter{equation}{0}
\setcounter{figure}{0}
\setcounter{table}{0}
\setcounter{page}{1}
\makeatletter
\renewcommand{\theequation}{S\arabic{equation}}
\renewcommand{\thefigure}{S\arabic{figure}}
\renewcommand{\bibnumfmt}[1]{[S#1]}
\renewcommand{\citenumfont}[1]{S#1}

\section{Mixture preparation}

The two-component BEC is prepared by employing the technique described in~\cite{Farolfi20,Farolfi21}, where we coherently transfer half of the atomic population to the $\ket{F,m_F}=
\ket{1,1}$ state through an adiabatic rapid passage (ARP) sequence.
In the presence of a uniform magnetic field $B=1.3$ G oriented along the vertical direction, we apply a microwave coupling field that addresses the two-photon transition with an effective Rabi frequency $\Omega_R = 2\pi\times \SI{180}{Hz}$ initially red-detuned from the transition to the $\ket{1,1}$ state. The detuning is then slowly reduced to zero by keeping the microwave frequencies constant and by modifying the Zeeman splitting with a $\SI{40}{\ms}$-long magnetic field ramp. This leads to an adiabatic rotation of the internal state from $\ket{1,-1}$ to a coherent superposition of  $\ket{1,-1}$ and $\ket{1,1}$ homogeneous along the cloud \cite{SFarolfi21}. Stabilization of the mixture against spin-changing collisions is ensured by an off-resonant additional microwave field with Rabi frequency \SI{6.88}{kHz} between $\ket{1,0}$ and $\ket{2,0}$, which introduces an artificial quadratic Zeeman shift~\cite{SBienaime16}.

\section{Measurement of $\Omega_R$}

We introduce a coherent Rabi coupling between the states $\ket{1, -1}$ and $\ket{1, +1}$ using a two-photon transition~\cite{SFarolfi21}, by means of two microwave radiations far detuned from the transition $\ket{1, \pm 1} \rightarrow \ket{2, 0}$ with Rabi frequency of \SI{5.4}{\kHz}. The strenght of the two-photon coupling is modified by changing the detuning from the single-photon transitions.
The measurement of the strength of the two-photon Rabi coupling is performed inducing on-resonance Rabi oscillations on a thermal cloud of atoms initially in $\ket{-1}$. The coupling is applied for a variable time $t$ and the normalized relative population of the two states, $Z = \frac{n_+-n_-}{n_+ + n_-}$, is measured. The two-photon resonance condition has been calibrated with a spectroscopic measurement as discussed in \cite{SFarolfi21}. The use of thermal clouds is motivated by the necessity to avoid mean field effects on the measurement.
\begin{figure}[h]
    \centering
    \includegraphics[width = \columnwidth]{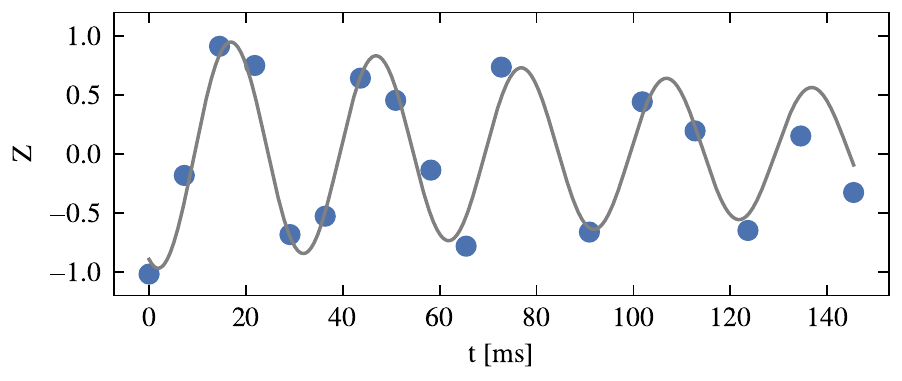}
    \caption{Measurement of the relative population as a function of time for Rabi coupling strength used in the data of Fig. 4 of the main text. The data are fitted with an exponentially damped sine.}
    \label{fig:SMfig1}
\end{figure}
A damped sine fitted on the data shown in Fig.\ref{fig:SMfig1} provides a two-photon Rabi frequency $\Omega_R = 2\pi \times \SI{33\pm 1}{Hz}$ [data in Fig.~4(a)]. A similar procedure leads to the value of $2\pi \times$ \SI{80 \pm 2}{Hz} for data in Fig.~4(b).
Small deviations are due to shot-to-shot noise in the bias field, that remains below 10 Hz thanks to the magnetic shield in our apparatus~\cite{SFarolfi19}.
The coherence of the oscillation has a $1/e$ lifetime of about $\SI{150}{ms}$, compatible with the expected coherence lifetime due to atom losses, mainly induced by the coupling through the lossy state $\ket{2, 0}$.

\section{Calibration of $\omega_p$}

For the data sets shown in the main text, the determination of $\mu_d$ and $\mu_s$ experimentally follows the procedure demonstrated in \cite{SFarolfi21}, where the frequency of small oscillations of $Z$ around the equilibrium point (plasma oscillations) is measured at the center of the condensate.
To perform the measurement, the system is prepared in the equilibrium point with the ARP procedure described above, then the phase of the coupling field is suddenly changed by $0.1\pi$, triggering the oscillation. After a variable time $t$, we image the population in each state and measure the frequency of oscillation 
\begin{equation}
 \omega_p=\sqrt{\Omega_R(\Omega_R+2\mu_s/\hbar)}
\end{equation}
of the relative population, from which we extract $\mu_s$.
From the measured $\mu_s$, we calculate $\mu_d$ as:
\begin{equation}
    \mu_d = \frac{3}{4}\left(\frac{g+g_{12}}{g-g_{12}}\right) \mu_s
\end{equation}
which stems from Eq.\,\ref{neffdens}-\ref{neffspin}.

\begin{figure}[h]
    \centering
    \includegraphics[width = \columnwidth]{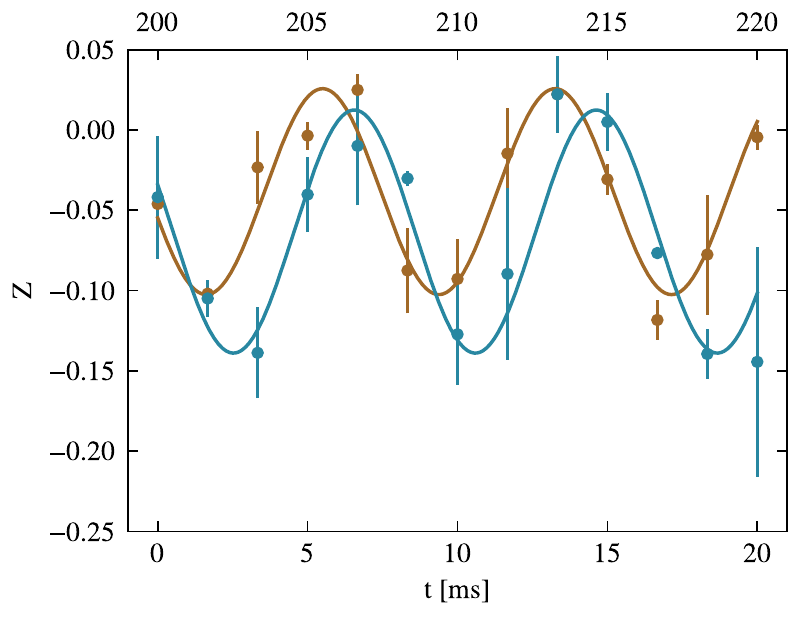}
    \caption{Measurement of the plasma frequency $\omega_p$ in the center of the cloud, with applied coupling $\Omega_R = 2\pi\times\SI{33}{Hz}$. Data in red have been taken right after the mixture preparation (lower horizontal axis), while light blue points have been measured 200 ms after the initial quench in $\Omega_R$ (upper horizontal axis). }
    \label{fig:SMfig2}
\end{figure}

Figure \ref{fig:SMfig2} shows the calibration of $\omega_p$ relative to the data of Fig.\,4 in the main text. The quench on the phase of the microwave coupling field is performed right after the end of the ARP preparation (red) and we measure $\omega_p = 2\pi \times \SI{129 \pm 3}{Hz}$.
To quantify the effect of atomic losses due to the presence of the coupling, we repeat the measurement by delaying the quench by \SI{200}{ms} (blue), which corresponds to the longest time for which the coupling is applied.
The measured value of $\omega_p = 2\pi\times \SI{124 \pm 3}{Hz}$ confirms that the effect of atomic losses is small. This small change is taken into account in the dispersion relations of Fig.\,4.

Furthermore, the modulation itself introduces additional losses. These are taken into account by introducing a rescaling factor in the value of $\mu_s$ extracted from the plasma frequency by looking at the atom number at the end of the modulation. 

\section{Role of the mixture balance}
\begin{figure}[h]
    \centering
    \includegraphics[width = \columnwidth]{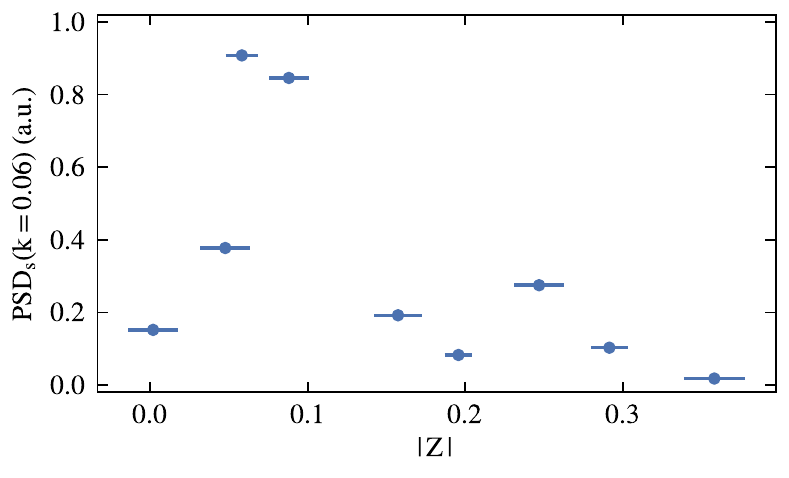}
    \caption{Amplitude of the dominant spin mode for different population imbalance. The maximum amplitude of the spin mode  is at a small but finite unbalance due to the increased coupling to the density channel where the modulation happens. For very large unbalance, the visibility drops.
    In this dataset, the experimental parameters are $\omega_M/2\pi = \SI{200}{Hz}$, modulation time $\SI{300}{ms}$, $\alpha = 0.6$. }
    \label{fig:SMfig3}
\end{figure}

In our two-component system, the spin and density excitation modes are exactly independent only for equal densities of each components $n_+ = n_-$, while the modes become hybridized in case of unequal populations \cite{SAbad2013}.
We investigate the role of unequal populations on the visibility of the Faraday pattern. To do so, we apply the experimental protocol on mixtures that are prepared with different imbalances, then we extract the amplitude of the dominant $k$ in the spin mode as a function of the imbalance, see \aref{fig:SMfig3}.
Starting from equal populations, the visibility of the pattern grows for finite unbalance at about 0.07. This is expected since we modulate the density of the mixture and by hybridizing the modes the coupling between the spin and density channels increases.
For larger unbalance, the visibility of the spin channel decreases, eventually reaching zero for a fully polarized mixture.

\section{Mathieu equation}

In this Section we focus on the origin of the weaker signals in Fig. 3 and Fig. 4 of the main text, where one can clearly see additional features at $\omega_M=2\omega_{s,d}(k)/2$ and even at $\omega_M=2\omega_{s,d}(k)/3$ (dashed lines).\\
A physical explanation for such an observation can be obtained by modelling the parametric emission considering an uniform one-dimensional condensate, whose chemical potential is $\mu=g n$, with $g$ the interaction strength and $n$ the density.
The effect of the modulation applied in the experiment, can be modeled by a modulation of the effective coupling constant, $g(t) = g[1+f(t)]$, with $f(t)=\eta \sin(\omega_M t)$ and $|\eta|\ll 1$. 

Within Bogoljubov perturbation theory, the dynamics of the excitations is described by a time dependent Hamiltonian, which, apart from irrelevant constant terms and up to quadratic order reads: 
\begin{equation}
\begin{split}
    \ham(t) &= \sum_{ k\ne 0} \big[\varepsilon(k)+F_k(t)\big] \op b_{ k}^\dagger \op b_{ k} + \sum_{k> 0} F_k(t)  \Big(\op b_{ k}^\dagger \op b_{- k}^\dagger + h.c. \Big)
\end{split}
\label{ham}
\end{equation}
where $\hat{b}_{k}$ ($\hat{b}_{k}^\dagger$) annihilates (creates) a quasi-particle of momentum $k$ and $\epsilon(k)=\hbar\omega(k)$ is the quasi-particle energy. The term $F(k,t) = \mu S(k)f(t)$ is proportional to static structure factor $S(k)$ of the BEC, which is related to the probability that a density probe  transfers a momentum $k$ to the system. According to Eq.(\ref{ham}), the periodic modulation generates pairs of counter-propagating quasi-particles of momenta $\pm k$ out of the Bogoljubov vacuum.

The Heisenberg equations for the operators $b$ and  $b^\dagger$ read:
\begin{equation}
    i\hbar\partial_t \binom{\op b_k}{\op b_{-k}^\dagger } = \begin{pmatrix}
    \varepsilon_k(t) & F_k(t) \\ -F_k(t) & -\varepsilon_k(t)
    \end{pmatrix} \binom{\op b_k}{\op b_{-k}^\dagger }
    \label{heis_eq}
\end{equation}
Let us now consider a solution of the form:
\begin{equation}
     \binom{\op b_k(t)}{\op b_{-k}^\dagger(t) } = \begin{pmatrix}
    \beta_1(t) & \beta_2(t) \\ \beta_2^*(t) & \beta_1^*(t) 
    \end{pmatrix} \binom{\op b_k(0)}{\op b_{-k}^\dagger(0) }
    \label{ansatz}
\end{equation}
where $\beta_1, \beta_2$ are some time-dependent amplitudes with initial conditions $\beta_1(0)=1$ and $\beta_1(0)=0$.
Solving System \ref{heis_eq} with Ansatz \ref{ansatz} leads to:
\begin{equation}
    \hbar^2\partial_t^2 \zeta + \big[\varepsilon^2_k(t)-F^2_k(t)\big] \zeta= 0 
    \label{eq:beta_eq}
\end{equation}
where $\zeta = \beta + \beta^*$ is the real part of the amplitudes. The index $i=1,2$ was removed, being the equation of motion the same for both parameters. 
Finally, if we define a dimensionless time $\tau = \omega_M t/2$,  \aref{eq:beta_eq} can be cast in the form of a Mathieu equation \cite{Bukov12}:
\begin{equation}
    \partial_\tau^2 \zeta + A(k,\omega) \Big[ 1 + B(k) \sin(2\tau) \Big] \zeta = 0
    \label{eq:mathieu}
\end{equation}
where the dimensionless coefficients $A,B$ are given by:
\begin{align}
    A(k, \omega) = \frac{\omega^2(k)}{(\omega_M/2)^2} \qquad 
    B(k) 
    = 2\eta \frac{\mu}{\varepsilon(k)} S(k)
\end{align}
The analysis of Mathieu equations [\aref{eq:mathieu}], for $B\ll 1$ (that is $\eta \ll 1$), reveals the presence of a series of instability lobes centred around momenta satisfying $A(k_l,\omega)=l^2$, with integer $l$, namely $\omega(k_l) = l\omega_M/2$. The larger $l$ the smaller the  region and the instability amplification coefficient (see, e.g., \cite{SLandau1935}), therefore the smaller the visibility for the same modulation amplitude.

\begin{figure}
    \centering
    \includegraphics[width = 0.8\linewidth]{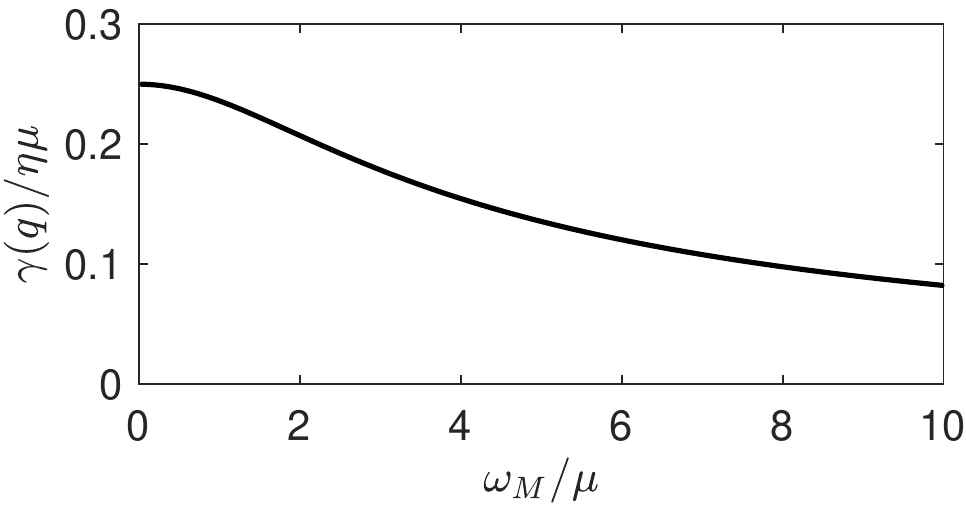}
    \caption{Instability rate of the mostly excited mode, according to Mathieu theory (formula \ref{gammaMathieu}). }
    \label{fig:gamma}
\end{figure}

The largest lobe is thus the one associated to $l=1$ and the mostly excited momentum $q$ is defined by $\omega(q)= \omega_M/2$. The instability rate, measured with respect to the dimensionless time $\tau$, associated to such mode is:
\begin{equation}
    \gamma(q) = \frac{\eta\mu S(q)}{2\omega(q)} = \frac{\eta\mu S(q)}{\omega_M} 
    \label{gammaMathieu}
\end{equation}
As shown in figure \ref{fig:gamma}, this quantity is constant at small momenta, $\gamma(q\to 0) \sim \eta\mu/4$ and decreases as $\sim \mu/\omega_M$ for $\omega_M/2 \gtrsim \mu$.

In the case of an unpolarized BEC mixture, spin and density degrees of freedom can be decoupled, and the Hamiltonian can be written as the sum of a spin and a density hamiltonian, $\ham_{d,s}$, which have the same form of Eq.\ref{ham}: therefore the above derivation of the Mathieu equation holds for the BEC mixture if we replace $\mu$ and $\varepsilon(k)$ with the effective chemical potentials $\mu_{d,s}$ and dispersion relations $\varepsilon_{d,s}(k)$ for the spin and density channels, and we insert the proper structure factors:
\begin{equation}
    S_d(k) = \frac{\hbar k^2/2m}{\omega_d(k)} \qquad S_s(k) = \frac{\hbar k^2/2m + \Omega_R}{\omega_s(k)}
\end{equation}

\section{Modulation frequency spectrum}
The signal appearing for $\omega_M \approx \omega_p$ in Fig.~4 of the main text can be attributed either to resonances of the Mathieu equation at $\omega_M=2 \omega(k)/l$, with $l = 2$ or,
trivially, to an anharmonic modulation of $\omega_{\perp}$. In the latter case, higher harmonics of the modulation frequency might be introduced by non-linearities in the system and $\omega_{\perp}(t)=\omega_{\perp}(0)[1+\alpha\sin(\omega_{M}t) + \alpha_1\sin(2\omega_{M}t) + ...]$.
We measure in time the intensity $I$ of the trapping laser and extract the harmonic components $\alpha$, $\alpha_1$ of $\omega_{\perp}(t)$ from its Fourier spectrum. 
The outcome of this analysis shows that $\alpha_1 \simeq \alpha/20$.
We repeat the measurement of Fig.~4 specifically at $\omega_M = 2\omega_p$ with a modulation amplitude reduced to 1/10 of its original value. In this way, we directly address the effect of the second harmonic component. With this protocol, we do not observe any spin excitation, therefore the role of the higher harmonics of the driving is negligible in our measurement. 

\section{Geometrical Reduction}

In order to determine the effective dispersion relation for longitudinal modes, it is convenient to develop a more accurate theoretical description. Let us then consider a three-dimensional BEC mixture in a symmetric configuration: mass and intracomponent interactions do not depend on the spin state, and all atoms are subject to the same trapping potential. The set of Gross-Pitaevskii (GP) equations describing the dynamics of the two components, hereafter labeled $\pm$, reads: 
\begin{equation}
\begin{split}
    i \hbar \partial{\psi_\pm}{t} &= \left(-\frac{\hbar^2\nabla^2}{2m}  + V + g|\psi_\pm|^2 + g_{12} |\psi_\mp|^2 \right)\psi_\pm \\
    &\qquad -  \frac{\hbar \Omega_R}{2} \psi_\mp
\end{split}  
\label{GPE}
\end{equation}
where $g$ ($g_{12}$) is the interspecies (intraspecies) interaction strength, $\Omega_R \ge 0$ is the Rabi frequency and $V$ describes an highly asymmetric trap, such that $\omega_\perp \equiv \omega_y = \omega_z \gg \omega_x$. To simplify calculations and focus only on the role of the transverse confinement, let us assume $\omega_x = 0$.
For suitable choices of the parameters the ground state of the system is unpolarized and well approximated by the Thomas-Fermi (TF) solution
\begin{equation}
    (g+g_{12}) \frac{n( r)}{2} = \mu +  \frac{\hbar \Omega_R}{2} -\frac 1 2  m\omega_\perp^2 r^2
\end{equation}
where $r = \sqrt{y^2+z^2}$ is the transverse coordinate, $n(r)$ is the total density and the chemical potential $\mu$ is fixed by the normalization condition for the order parameter. With these assumptions the condensed cloud has the shape of a long and thin cylinder of transverse radius $R = \sqrt{(2\mu+\hbar\Omega_R)/m\omega_\perp^2}$. 

Given a perturbation, such that $|\psi_\pm|^2 = n/2 + \delta n_\pm$ and arg$(\psi_\pm) = -\mu t/ \hbar + \delta S_\pm$, it is convenient to define spin and density variables as $\delta n_{d,s} = \delta n_+ \pm \delta n_-$ and $\delta S_{d,s}  = \delta S_+ \pm \delta S_-$. 
By linearizing the GP equations \ref{GPE} to first order in $\delta n_{d,s}, \delta S_{d,s}$, one can write hydrodynamic-like equations: 
\begin{align}
&\begin{cases}
	\partial_t \, \delta n_d + \cfrac{\hbar}{2m}\nabla (  n \nabla \delta S_d ) = 0\\
	\hbar \partial_t\, \delta S_d + (g+g_{12}) n_d + P_d = 0\\
\end{cases} \label{hydro_den}\\
&\begin{cases}
	\partial_t \,\delta n_s + \cfrac{\hbar}{2m} \nabla ( n \nabla \delta S_s ) =  n \Omega_R \delta S_s\\
	\hbar \partial_t \,\delta S_s + \left(g-g_{12}+ \cfrac{\hbar\Omega_R}{n} \right) \delta n_s + P_s = 0
\end{cases}\label{hydro_spin}
\end{align}
where $P_{d,s}$ are corrections to the hydrodynamic regime, coming from second order derivatives of the TF profile:
\begin{equation}
P_{d,s} = -\frac{\hbar^2}{2m n} \vec\nabla\left[n \vec\nabla\left(\frac{\delta n_{d,s}}{n}\right)\right]\label{qpressure}
\end{equation}
Since we are interested in longitudinal modes, we look for solutions of the form $\delta n_{d,s}(\vec{r}) = f(x) w(r)$.
Assume the healing length of the perturbation $\xi$ is much smaller than the transverse size of the condensate, so that the terms $P_{d,s}$ are negligible: this will always be true for density perturbations, as well as in the case of a single component condensate; it will also be valid for spin perturbations in the limit $g_{12} \ll g$, when $\xi_s \gtrsim \xi_d$. Let us also start, for simplicity with the case $|\Omega|=0$. 
Under these conditions equations \ref{hydro_den}, \ref{hydro_spin} can be cast in the form:
\begin{align}
\partial_t^2\, \delta n_{d,s} &= \frac{(g\pm g_{12})}{2 m} \nabla \big( n \nabla \delta n_{d,s} \big) \label{hydro_norabi}
\end{align}
In order to solve them we need to choose an Ansatz for the transverse profile $w(r)$ and integrate out the coordinate $r$. As long as $\xi \ll R$, an approximate solution for the perturbation profile can be found in the TF approximation, $\delta \mu_\pm = g\delta n_\pm + g_{12} \delta n_\mp$,
which gives:
\begin{equation}
    \delta n_{d,s}= \delta n_+ \pm \delta n_- = \frac{\delta\mu_+ \pm \delta\mu_-}{g+g_{12}} \label{eq_norabi}
\end{equation}
where $\delta\mu_\pm$ are generic variations of the two chemical potentials. Equation \ref{eq_norabi} suggests that the correct Ansatz for the transverse profile of the perturbation if $\xi \ll R$ is a uniform one, $w(r) = \theta(R-r)$. Equation \ref{hydro_norabi} becomes:
\begin{equation}
\partial_t^2 f =  \frac{(g\pm g_{12})}{2m} n(r)  \partial_x^2 f 
\end{equation}
Integration on both sides in the domain $r\le R$ leads to:
\begin{equation}
\partial_t^2 f =  \frac{(g\pm g_{12})}{2m} n_{\text{eff},d,s}\, \partial_x^2 f \label{waveeq}
\end{equation}
where $n_{\text{eff},d,s}=n_0/2$, $n_0 \equiv n(r=0)$ being the peak density. If $f$ is a plane-wave with frequency $\omega$ and momentum $k$, equation \ref{waveeq} gives a linear dispersion with speed of sound $c_{d,s} = \sqrt{(g\pm g_{12})n_{\text{eff},d,s}/2m}$. This renormalization factor for the speed of sound in a trapped condensate is a well known result for single component systems.
However BEC mixtures admit a richer phenomenology. For instance, spin modes can be made soft enough to violate the hydrodynamic regime, that is, $\xi_s \sim R$: if this is the case, Eq.(\ref{eq_norabi}) does not hold. 
In what follows we prove that, the solution $w(r)=n(r)$ allows to straightforwardly include corrections to the hydrodynamics results, both with and without Rabi coupling. Hence it is well suited to describe spin perturbations outside the TF regime.
If we include the correction term $P_s$, the system of equations \ref{hydro_spin} can be cast in the form:
\begin{equation*}
\begin{split}
\partial_t^2 \delta n_s &= \frac{1}{m} \vec\nabla \left\{ \frac{n}{2} \vec\nabla\left[\left(g-g_{12}+\frac{\hbar \Omega_R}{n} \right) \delta n_s + P_s\right]\right\} \\
&\quad - n \frac{\Omega_R}{\hbar} \left[\left(g-g_{12}+\frac{\hbar \Omega_R}{n} \right) \delta n_s + P_s \right] 
\end{split}
\end{equation*}
The assumption $w(r)=n(r)$ leads to:
\begin{equation}
\begin{split}
    n \partial_t^2 f &= -\frac{n}{\hbar^2} \left(- \frac{\hbar^2 D_x^2}{2m}  + (g-g_{12})n \right) \frac{\hbar^2 D_x^2}{2m} f \\
    & \qquad + \frac{(g-g_{12})}{2m}  \vec\nabla \big(n \vec\nabla n\big) f
\end{split}
\end{equation}
where $D^2_x = \partial_x^2 - 2m\Omega_R/\hbar$ is a modified second derivative along $x$, which coincides with the standard partial derivative in the absence of Rabi coupling.
When integrated along the transverse direction, the gradient term proportional to $f$ vanishes because the density $n(r)$ vanishes at the boundaries of the integration domain. The remaining terms, when integrated with respect to $r$, give:
\begin{equation}
\partial_t^2 f = \left(- \frac{\hbar \partial_x^2}{2m} + \frac{(g-g_{12})n_{\text{eff,s}}}{\hbar} + \Omega_R \right) \left(\frac{\hbar \partial^2_x}{2m} + \Omega_R \right)   f
\label{rabi_waveeq}
\end{equation}
where the density is renormalized as $n_\text{eff,s} = 2n_0/3$. In case $f$ is a plane wave of momentum $k$, equation \ref{rabi_waveeq} translates in the usual Bogoljubov dispersion relation for spin modes with modified speed of sound and mass.

Summarizing, our calculations show that longitudinal excitations over a cigar-shaped trapped BEC mixture are Bogoljubov modes satisfying the usual dispersion relations in the density and spin channels:
\begin{align}
    \omega_d(k) &= \sqrt{\frac{\hbar k^2}{2m} \left(\frac{\hbar k^2}{2m} + \frac{2 \mu_d}{\hbar} \right)}\\
    \omega_s(k) &= \sqrt{\left(\frac{\hbar k^2}{2m} + \frac{2 \mu_s}{\hbar} + \Omega_R \right)\left(\frac{\hbar k^2}{2m}  +\Omega_R \right)}
    \label{dipersionTheory}
\end{align}
where $\mu_{d,s} = (g\pm g_{12})n_{\text{eff},d,s}/2$ are effective chemical potentials and the effective speed of sound and plasma frequency are given by:
\begin{align}
    c_d &= \sqrt{\frac{(g+g_{12})n_{\text{eff},d}}{2m}}= \sqrt{\frac{\mu_d}{m}} \label{effc_den}\\
    c_s &= \sqrt{\frac{(g-g_{12})n_{\text{eff},s}+\hbar \Omega_R}{2 m}} = \sqrt{\frac{\mu_s+\hbar \Omega_R/2}{m}} \label{effc_spin} \\
    \omega_p &= \sqrt{\Omega_R (\Omega_R + 2 \mu_s/\hbar)} \label{effplasma_spin}
\end{align}
The effective density depends on the parameters' values and can be different for density and spin modes. More specifically, density excitations will always satisfy: 
\begin{equation}
    n_{\text{eff},d} = n_0/2
    \label{neffdens}
\end{equation}
whereas for spin excitations in the presence of Rabi coupling one always finds: 
\begin{equation}
        n_{\text{eff},s} = 2n_0/3
    \label{neffspin}
\end{equation} 
In the case of spin modes in the absence of Rabi coupling, we can define upper and lower bounds to the effective density: $n_\text{eff,s}^< = n_0/2$, valid in the limit $g_{12}\ll g$, and $n_\text{eff,s}^> = 2n_0/3$, for $g_{12}\to g$, and estimate the actual one as follows:
\begin{equation}
	n_\text{eff,s} = \left(1-\frac{g_{12}}{g}\right) n_\text{eff,s}^< + \frac{g_{12}}{g} n_\text{eff,s}^>
\end{equation}
In particular, for sodium $g_{12} \sim 0.93 g$, so that the effective density in the experiment can be reduced to equation \ref{neffspin}.

\end{document}